\documentclass [aip,jcp,reprint]{revtex4-1}
\usepackage{graphicx}
\usepackage{hyperref}
\usepackage{float}
\usepackage{longtable}
\newcommand{\footnotelabel}[2]{
  \footnote{#2}
  \newcounter{#1}
  \setcounter{#1}{\value{footnote}}
}
\newcommand{\footnoteref}[1]{
  \footnotemark[\value{#1}]
}
\begin{document}
\title{Monovalent Ion Condensation at the Electrified Liquid/Liquid Interface}
%\date{\today}

\author{Nouamane Laanait}
\email{nlaana1@uic.edu}
\author{Jaesung Yoon}
\author{Binyang Hou}
\author{Mark L. Schlossman}
\email{schloss@uic.edu}
\affiliation{Department of Physics, University of Illinois, Chicago, Illinois 60607, USA}
\author{Petr Vanysek}
\affiliation{Department of Chemistry and Biochemistry, Northern Illinois University, DeKalb, Illinois 60115, USA}
\author{Mati Meron}
\author{Binhua Lin}
\affiliation{The Center for Advanced Radiation Sources, University of Chicago,Chicago, Illinois 60637, USA}
\author{Guangming Luo}
\affiliation{Division of Chemical Sciences and Engineering, Argonne National Laboratory,
Argonne, Illinois 60439, USA}
\author{Ilan Benjamin}
\affiliation{Department of Chemistry, University of California, Santa Cruz, California 95064, USA}

\begin{abstract}
X-ray reflectivity studies demonstrate the condensation of a monovalent ion at the electrified interface between electrolyte solutions of water and 1,2-dichloroethane.  Predictions of the ion distributions by standard Poisson-Boltzmann (Gouy-Chapman) theory are inconsistent with these data at higher applied interfacial electric potentials.  Calculations from a Poisson-Boltzmann equation that incorporates a non-monotonic ion-specific potential of mean force are in good agreement with the data.
\end{abstract}
\maketitle
\indent Interfacial ion distributions underlie numerous electrochemical and biological processes, including electron and ion transfer across charged biomembranes and energy storage in electrochemical capacitors. The solution to the Poisson-Boltzmann equation for a planar geometry, Gouy-Chapman theory, including modifications with a Stern layer, is often used to predict ion distributions near those interfaces.\cite{Gouy1910,*Chapman1913,*Stern1924}  We showed previously that the predictions of such theories are inconsistent with x-ray reflectivity measurements of ion distributions at an electrified liquid/liquid interface.\cite{Luo2006a,*Luo2006b}  Instead, an ion-specific Poisson-Boltzmann equation (PB-PMF) that incorporated a potential of mean force (PMF) for each ion produced excellent agreement with the x-ray results.\cite{Luo2006a,*Luo2006b}  The PB-PMF theory accounts for interactions and correlations between ions and solvents that are left out of Gouy-Chapman theory.  This approach has promise for understanding ion-specific effects that are relevant to many chemical processes.\cite{Lima2008a} \\
\begin{figure}[b]
\begin{center}
\includegraphics[scale=0.8]{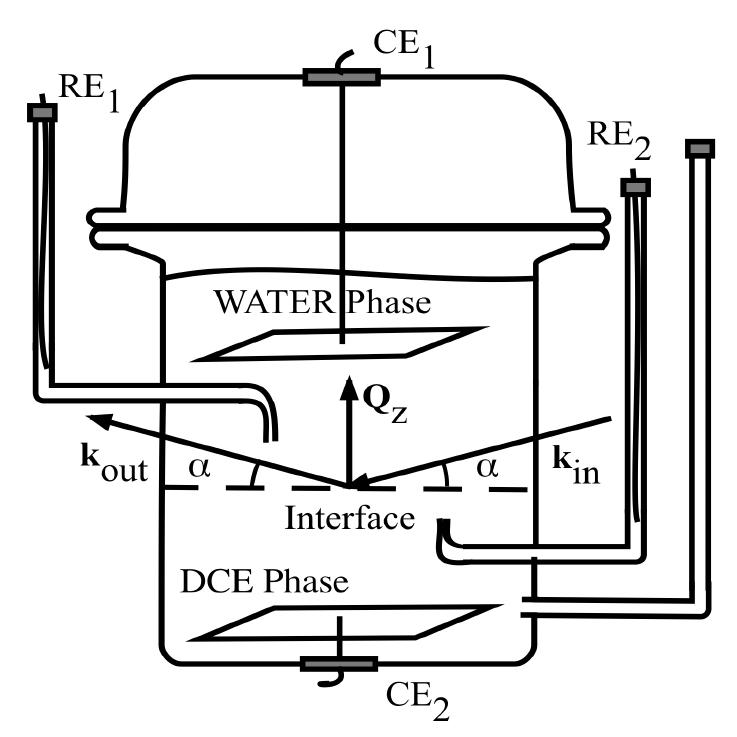}
\caption{Circular glass sample cell and x-ray kinematics. Electrochemical cell diagram:	Ag$\arrowvert$AgCl $\arrowvert$ 0.1M NaCl $\arrowvert$ water $\arrowvert$ +20 mM HEPES $\arrowvert$ 5 mM BTPPATPFB $\arrowvert$ DCE $\arrowvert$ 10 mM LiCl+1 mM BTPPACl $\arrowvert$ water $\arrowvert$ AgCl $\arrowvert$ Ag. A four-electrode potentiostat (Solartron 1287) is used to apply potential at counter electrodes (CE$_{1,2}$, 9 cm$^{2}$ Pt mesh) and monitor potential at reference electrodes (RE$_{1,2}$) in Luggin capillaries within 4 mm of interface. The liquid/liquid interface of 7 cm diameter is pinned by a Teflon strip (affixed to the glass wall) and flattened by adjusting the volume of DCE phase. Volume ratio of water:DCE is 2:1. The x-ray wave vector transfer $\vec{Q}=\vec{k}_{out}-\vec{k}_{in}$.}
\label{fig:cell}
\end{center}
\end{figure}
\indent Here, we demonstrate the condensation of a monovalent ion at a liquid/liquid interface. Recent theory proposes that condensation of multivalent ions is the result of strong ion-ion correlations.\cite{Grosberg2002}  However, these theories do not predict such distributions for monovalent ions. Our current results can be understood by PB-PMF theory.  We have chosen to fit our x-ray data to the potentials of mean force instead of fitting to a model of the electron density profile because a single PMF for each ion determines the ion distributions for all interfacial potentials.\\
\indent The system under study is the liquid/liquid interface between a 100 mM aqueous solution of NaCl (Fisher Scientific) including 20 mM HEPES to buffer the pH to 7.0, and a 5 mM solution of bis(triphenyl phosphoranylidene) ammonium tetrakis(pentafluorophenyl) borate (BTPPA$^{+}$, TPFB$^{-}$) in 1,2-dichloroethane (DCE, Fluka). Water was produced by a Barnstead Nanopure system and DCE was purified using a column of basic alumina.  BTPPATPFB was synthesized from BTPPACl (Aldrich) and LiTPFB (Boulder Scientific).\cite{Fermin1999a} Conductance measurements using the method in Ref. \cite{Raymond1949} determined that ~54\% of BTPPATPFB is dissociated in DCE.\\
\indent The electric potential difference $\Delta\phi^{w-o}(=\phi^{water} - \phi^{oil})$   between the water and oil (DCE) phases is given by the applied potential difference across the electrochemical cell (Fig.\ref{fig:cell}) minus the potential of zero charge ($\Delta\phi^{w-o}=\Delta\phi^{w-o}_{applied} - \Delta\phi^{w-o}_{pzc}$).  We determined   $\Delta\phi^{w-o}_{pzc}= 318\pm 3$ mV by measuring the interfacial tension as a function of $\Delta\phi^{w-o}_{applied} $\cite{Schmickler1996}. The ions Na$^{+}$ and Cl$^{-}$ stay primarily in the aqueous phase and BTPPA$^{+}$ and TPFB$^{-}$ stay in the DCE phase throughout the potential range studied. When $\Delta\phi^{w-o} \neq0$,  the ions form back-to-back electrical double layers at the interface.  For example, when $\Delta\phi^{w-o} > 0$, the concentrations of Na$^{+}$and TPFB$^{-}$ are enhanced at the interface while those of Cl$^{-}$ and BTPPA$^{+}$ are depleted.  The variation of ionic concentration along the interfacial normal  produces a variation in the electron density profile $\rho(z)$   (averaged over the x-y plane) that is probed by x-ray reflectivity.\\
\begin{figure}[H]
\begin{center}
\includegraphics[scale=0.8]{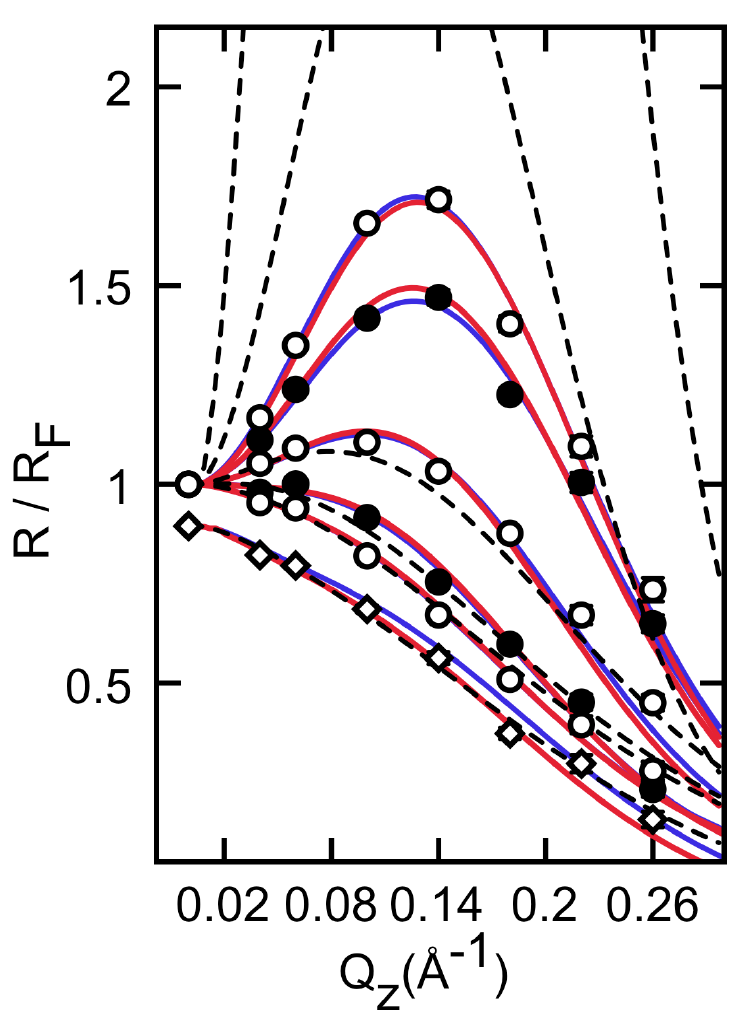}
\caption{X-ray reflectivity $R(Q_{z})$ normalized to Fresnel reflectivity $R_{F}(Q_{z})$ for various potentials across the water/1,2-dichloroethane interface as a function of wave vector transfer normal to the interface at T=296 K. From top to bottom $\Delta\phi^{w-o}=0.33$ V ($\circ$), $0.28$ V ($\bullet$), $0.18$ V($\circ$), $0.08$ V ($\bullet$), $-0.02$ V($\circ$), and $-0.12$ V($\diamond$)  (offset for viewing purposes). Dashed lines: Gouy-Chapman theory. Solid lines: PB-PMF. Red and blue lines indicate the use of two different PMFs for TPFB$^{-}$ (see text).}
\label{fig:RRf}
\end{center}
\end{figure}

\indent X-ray reflectivity measurements $R(Q_{z})$  from the electrified liquid/liquid interface were carried out at the ChemMatCARS sector of the Advanced Photon Source.\cite{Schlossman1997a} $R(Q_{z})$ is the reflected intensity normalized by the incident intensity (after subtraction of background scattering \cite{Zhang1999}) as a function of wave vector transfer $Q_{z}=(4\pi$/$\lambda)\sin\alpha$  , where $\lambda$ (=0.41255 $\pm$0.00005 \AA) is the x-ray wavelength and $\alpha$ is the angle of incidence (Fig. 1).  Figure 2  illustrates $R(Q_{z})$/$R_{F}(Q_{z})$ for different $\Delta\phi^{w-o}$. The variation of the peak amplitude in $R$/$R_{F}$ with increasing $\Delta\phi^{w-o}$ reveals the formation of a TPFB$^{-}$ layer, as discussed below.\\
\begin{figure}[tlc] 
\begin{center} 
\includegraphics[scale=0.8]{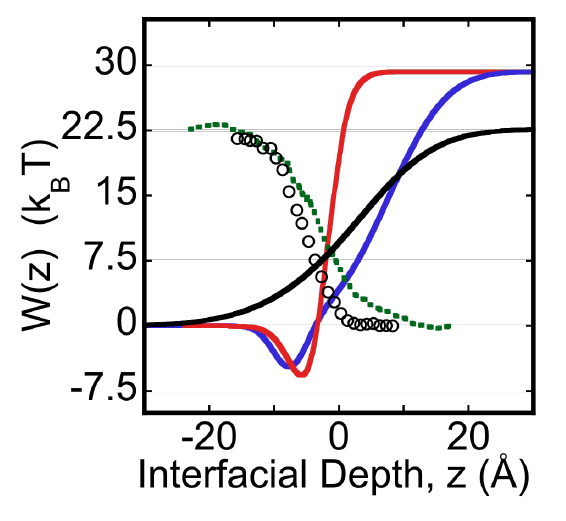}  
\caption{Potentials of mean force for BTPPA$^{+}$ (black) and TPFB$^{-}$ [$W_{TPFB^{-}}^{I}(z)$:red,$W_{TPFB^{-}}^{II}(z)$:blue] determined by fitting the reflectivity data in Fig. \ref{fig:RRf}. PMFs for Na$^{+}$ (green dots) and Cl$^{-}$ (circles) were calculated by MD simulations (see text) (Ref.\cite{Wick2008}.}
\label{fig:PMF}
\end{center} 
\end{figure}               

\indent We analyzed the x-ray data using the Poisson-Boltzmann (PB) equation,
\begin{equation}
\frac{d^{2}\phi(z)}{dz^{2}}=- \frac{1}{\varepsilon_{o}\varepsilon}\sum_{i}e_{i}c_{i}^{o} \exp[-\Delta E_{i}(z)/k_{B}T],
\label{eq:pb}
\end{equation}
which relates the electric potential $\phi(z)$ along the interfacial normal $z$ to the concentration profile of ion i, $c_{i}(z)=c_{i}^{o}\exp[-\Delta E_{i}(z)/k_{B}T]$  , with Boltzmann constant $k_{B}$, temperature $T$, charge $e_{i}$  of ion $i$  (BTPPA$^{+}$, TPFB$^{-}$, Na$^{+}$, and Cl$^{-}$), permittivity of free space $\varepsilon_{o}$, and dielectric constant $\varepsilon$ of either DCE (10.43) for $z<0$  or water (78.54) for $z>0$ . $\Delta E_{i}(z)$ is the energy of ion $i$  relative to its value in the bulk phase.  The bulk ion concentration $c_{i}^{o}$  is calculated from the Nernst equation\footnotelabel{gibbsenerg_footnote}{Gibbs energies of transfer: Na$^{+}$ (57 kJ/mol), Cl$^{-}$ (53 kJ/mol), BTPPA$^{+}$ (56 kJ/mol), TPFB$^{-}$ (72.5 kJ/mol), last two measured by partitioning
via UV-visible spectroscopy and mass spectroscopy}$^{,}$\cite{Volkov1996}.  Fitting to the data involves calculating the electron density $\rho(z)$ and $R/R_{F}$ from the ion concentration profiles $c_{i}(z)$ as described previously.\cite{Luo2006a,*Luo2006b}  For the purpose of calculating $\rho(z)$ from $c_{i}(z)$ the ions were modeled as spheres of diameter 2{\AA} for Na$^{+}$, 3.5{\AA}  for Cl$^{-}$, 12.6{\AA}  for BTPPA$^{+}$, and 10{\AA}  for TPFB$^{-}$, where the latter were estimated from the crystal structure of BTPPATPFB.\cite{Marcus1988}$^{,}$ \footnote{C.Zheng and P. Vanysek (unpublished)} The TPFB$^{-}$ ion provides the dominant ionic contribution to the electron density profile when $\Delta\phi^{w-o}>0$ . \\
\indent The Gouy-Chapman theory assumes that $E_{i}(z)=e_{i}\phi(z)$ in Eq. \ref{eq:pb}. Fits of $R/R_{F}$ to predictions of Gouy-Chapman theory (Fig. \ref{fig:RRf}, dashed lines) used only the interfacial roughness and a $Q_{z}$ offset (~$10^{-4}${\AA}$^{-1}$, a typical misalignment of the reflectometer) as fitting parameters. 
\begin{longtable*}{l *{6}{c}}
\multicolumn{7}{c}{\parbox{\LTcapwidth}{TABLE I. TPFB$^{-}$ PMF parameters obtained by fitting to the reflectivity data at $\Delta\phi^{w-o}=0.28$ and 0.33 V.}}\\
\\ 
\begin{ruledtabular} 
\begin{tabular}{l *{6}{c}} 
&$W(0)$ 	 & $L^{o}$	& $L^{w}$ 	& $z_{0}$	& $\sigma_{PMF}$	& $D$\\
&$(k_{B}T)$ & (\AA) & (\AA) & (\AA) & (\AA)  & $(k_{B}T)$ \\
\\
\hline
$W_{TPFB^{-}}^{I}(z)$ &$-5\pm0.5$ &$3\pm0.1$ & $9\pm4$ &$-3.5\pm0.2$ & $3.4\pm0.2$ & $-9\pm0.25$ \\
$W_{TPFB^{-}}^{II}(z)$ &$-25\pm0.5$ &$11\pm0.5$ &$10\pm2.7$ &$-7.5\pm0.3$ &$2.6\pm0.3$ &$-5.25\pm0.2$ \\ 
\label{table1}
\end{tabular} 
\end{ruledtabular} 
\end{longtable*}

These fits agree with the data at small $\Delta\phi^{w-o}$ ($-0.12$ V to $0.18$ V), but at larger $\Delta\phi^{w-o}$ (0.28 V and 0.33 V) $R/R_{F}$ is greatly overestimated  primarily because Gouy-Chapman theory predicts unphysically large ion concentrations near the interface. \\
 \indent Ion-specific effects can be included in Eq. \ref{eq:pb} by expressing $E_{i}(z)\approx e_{i}\phi(z)+W_{i}(z)$, where $W_{i}(z)$ is the potential of mean force (PMF) for each ion $i$.\cite{Luo2006a,*Luo2006b,Horinek2007,Daikhin2001} The PMF of Na$^{+}$ was calculated from a molecular dynamics (MD) simulation for a single ion.\cite{Benjamin2008} The PMF of Cl$^{-}$ was taken from an MD simulation in the literature.\cite{Wick2008} Fig. \ref{fig:PMF} illustrates the monotonic variation of $W_{i}(z)$ for Na$^{+}$ and Cl$^{-}$.  Due to the computational difficulties of simulating $W_{i}(z)$ for large molecular ions such as BTPPA$^{+}$ and TPFB$^{-}$ we used a phenomenological PMF previously introduced in Ref.\cite{Luo2006a,*Luo2006b},
\begin{equation}
W_{i}(z)=(W_{i}(0)-W_{i}^{p})\frac{\mathrm{erf}[|z|-\delta_{i}^{p}/L_{i}^{p}]}{\mathrm{erfc}[-\delta_{i}^{p}/L_{i}^{p}]}+W_{i}^{p},
\label{eq:erf_pmf}
\end{equation}
where $p(= w, o)$ refers to either the water phase $(z\geq0 )$ or the oil phase (DCE,$z\leq0$), $W_{i}^{o}-W_{i}^{w}$  is the Gibbs energy of transfer of ion $i$ from water to oil, $\delta_{i}^{p}$ is an offset to ensure continuity of $W_{i}(z)$ at $z = 0$, and $L_{i}^{p}$ characterizes the decay of $W_{i}(z = 0)$ to its bulk values $W_{i}^{w}$ and$W_{i}^{o}$. We used this monotonic PMF for BTPPA$^{+}$, but had to modify it for TPFB$^{-}$, as described below.  Since $W_{i}^{o}-W_{i}^{w}$ for BTPPA$^{+}$ is known (Ref. \footnoteref{gibbsenerg_footnote}) , the PMF of BTPPA$^{+}$ is characterized by 3 parameters determined by fitting to $R/R_{F}$ data at  $\Delta\phi^{w-o}= -0.12$ V where it is expected that the BTPPA$^{+}$ interfacial concentration is enhanced: $L^{w}_{BTPPA^{+}} (=14 +12/-6${\AA}), $L^{o}_{BTPPA^{+}} (=20 +11/-6${\AA}), and $W_{BTPPA^{+}}(0) (=13\pm 2 k_{B}T )$.  The large error bars on the PMF of BTPPA$^{+}$ are due to the small magnitude of the most negative $\Delta\phi^{w-o}$ that we studied.
\begin{figure}[b] 
\begin{center} 
 \includegraphics[scale=0.95]{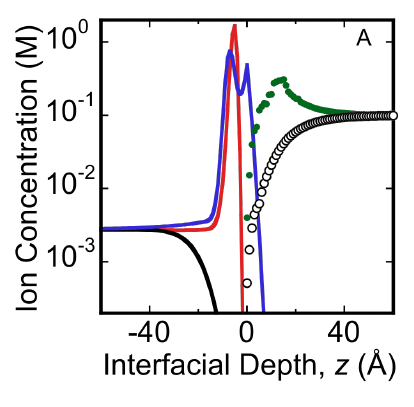}
 \includegraphics[scale=1]{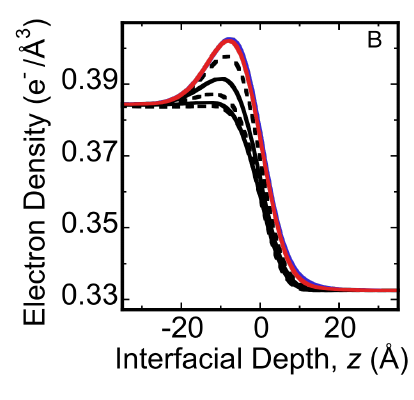}
\caption{(left) Ion concentration profiles (in units of molarity) at $\Delta\phi^{w-o}=0.33$V calculated from PB-PMF. BTPPA$^{+}$ (black), TPFB$^{-}$ [$W_{TPFB^{-}}^{I}(z)$ :red,$W_{TPFB^{-}}^{II}(z)$ :blue], Na$^{+}$ (dots) and Cl$^{-}$ (circles). (right) Electron density profiles for various potentials
calculated from PB-PMF. Top to bottom:  $\Delta\phi^{w-o}=0.33$V [$W_{TPFB^{-}}^{I}(z)$:red,$W_{TPFB^{-}}^{II}(z)$:blue], 0.28V (dashed), 0.18V (solid), 0.08(dashed), $-0.02$V (solid), and $-0.12$ V (dashed).}
\label{fig:iondens} 
\end{center} 
\end{figure}        

The x-ray reflectivity at the two highest positive potentials cannot be fit if Eq. \ref{eq:erf_pmf} is used to model the PMF for TPFB$^{-}$.  The simplest model that will produce the peaks in Fig. \ref{fig:PMF} is a single layer of TPFB$^{-}$ ions at the interface (note that a layer of Na$^{+}$, whose concentration is also enhanced at the interface, cannot provide the x-ray contrast required to fit the data). The TPFB$^{-}$ layer is modeled by an attractive well in the PMF. $W_{TPFB^{-}}(z)$  is given by Eq. \ref{eq:erf_pmf} plus a Gaussian function $D\exp[-(z-z_{0})^{2}/2\sigma^{2}_{PMF}$] for $z<0$  along with a constant offset at $z=0$ to maintain continuity (see Fig.\ref{fig:PMF}). Analysis with a Lorentzian produced similar results.\\
\indent The six parameters of $W_{TPFB^{-}}(z)$ [$z_{0}$, $D$, $\sigma_{PMF}$,  $L^{w}_{TPFB^{-}}$, $L^{o}_{TPFB^{-}}$, $W_{TPFB^{-}}(0)$] along with the $Q_{z}$ offset and the interfacial roughness ( $4.3 \textrm{\AA} < \sigma<5.1\textrm{\AA}$) are determined by fitting $R/R_{F}$ measured at   $\Delta\phi^{w-o}= 0.28$ V and 0.33 V, where the concentration of TPFB$^{-}$ is enhanced at the interface (Table \ref{table1}).  This fitting is performed under the constraint that the resultant $W_{ion}(z)$  produces $R/R_{F}$ in agreement with the data over the \textit{entire} range of potentials.  In addition, fitted PMFs were rejected if the fit value of the roughness $\sigma$ was unphysically small. In those cases an interfacial bending modulus\cite{Safran1994} on the order of $1000 k_{B}T$ would have been required to reconcile the discrepancy of $\sigma$ with its value predicted by capillary wave theory\cite{Luo2006}. In the case of the TPFB$^{-}$ PMF, two local minima in $\chi^{2}$-space (denoted $W_{TPFB^{-}}^{I}(z)$  and $W_{TPFB^{-}}^{II}(z)$ ) were found to satisfy these conditions.  Potential profiles that are intermediate between $W_{TPFB^{-}}^{I}(z)$ and $W_{TPFB^{-}}^{II(z)}$ do not satisfy these conditions. Most of these fits had values of $\sigma$ within one standard deviation of capillary wave theory predictions using the measured potential-dependent interfacial tension.\cite{Luo2006a,Luo2006b}. Fits to $W_{TPFB^{-}}^{II}(z)$ at  $\Delta\phi^{w-o}= 0.28$ V and 0.33 V had values of $\sigma$ within two standard deviations of capillary wave theory.\\
\indent The PB-PMF model with the $W_{i}(z)$ shown in Fig. \ref{fig:PMF} produces $R/R_{F}$ in good agreement with the data over the entire range of measured potentials (Fig. \ref{fig:RRf}).  The attractive wells for $W_{TPFB^{-}}^{I,II}$  have comparable depths (6 $k_{B}T$ for $W_{TPFB^{-}}^{I}(z)$ and 5 $k_{B}$T for $W_{TPFB^{-}}^{II}(z)$), FWHM, and centers (Table 1).  The ion concentration profiles $c_{i}(z)$, are calculated from Eq. \ref{eq:pb} using $W_{i}(z)$. Figure \ref{fig:iondens} shows that the $c_{i}(z)$ at the highest potential, $\Delta\phi^{w-o}= 0.33$ V, take the form of two back-to-back double layers with a sharply defined layer of TPFB$^{-}$.  The different $c_{i}(z)$ calculated from $W_{TPFB^{-}}^{I}(z)$  or $W_{TPFB^{-}}^{II}(z)$  differ mainly in the broadness of the profile, which in the case of  $W_{TPFB^{-}}^{I}(z)$ returns to its bulk value at  $z=0$, while $W_{TPFB^{-}}^{II}(z)$  allows TPFB$^{-}$ to penetrate slightly more into the water phase. The electron density profiles $\rho(z)$ calculated from the different $c_{i}(z)$ are almost identical (Fig. \ref{fig:PMF}), which demonstrates why our data cannot discriminate between $W_{TPFB^{-}}^{I}(z)$ and $W_{TPFB^{-}}^{II}(z)$  . \\
\indent The maximum density of TPFB$^{-}$ near the interface occurs at  $\Delta\phi^{w-o}= 0.33$ V and is 1 nm$^{2}$ per TPFB$^{-}$ ion when $W_{TPFB^{-}}^{I}(z)$   is used or 1.5 nm$^{2}$ per TPFB$^{-}$ ion when $W_{TPFB^{-}}^{II}(z)$  is used.  Both values represent a high-density layer for an ion of 1 nm diameter.  Although dense ionic layers have been observed in the interfacial adsorption of charged amphiphiles, \cite{Leveiller1991} the absence of a dense TPFB$^{-}$ layer at $\Delta\phi^{w-o}\approx0$ indicates that TPFB$^{-}$ is, at most, weakly amphiphilic. \\
\indent Simulations and supporting spectroscopy experiments indicate that highly polarizable ions (such as I$^{-}$ with a polarizability of  ~7.4{\AA}$^{3}$) are preferentially adsorbed to the water/vapor interface, though dense layers are not expected or observed.\cite{Winter2006,*Dang2002} We calculated the polarizability of TPFB$^{-}$ to be 42.9{\AA}$^{3}$.\cite{g03}$^{,}$\footnote{TPFB$^{-}$ geometry optimized at B3LYP/6-311++G(d,p) level with tightest constraints on convergence (opt=very tight, int=ultra fine). Resultant geometry agreed closely with crystal structure. Polarizability tensor calculated from optimized structure at the same level of theory.} This large polarizability may play a role in forming a dense layer at high potentials.  Also, Borukhov \textit{et al.} suggested that the entropy of the solvent can stabilize large ion adsorption.\cite{Borukhov1997} Additional theoretical work is required to determine the relevance of these two effects for the data presented here.\\
\indent The MD simulations of the potentials of mean force that we used for Na$^{+}$ and Cl$^{-}$ do not account for ion-ion correlations, but they do include ion-solvent and solvent-solvent correlations.  Such correlations also account for the monotonic form of $W_{BTPPA^{+}}(z)$ . However, as a result of modeling the x-ray reflectivity, the phenomenological $W_{TPFB^{-}}(z)$ in Fig. \ref{fig:PMF} must implicitly account for ion-ion correlations if they are important for the observed condensation. The description of this monovalent ion condensation within PB-PMF theory illustrates the utility of this approach in describing ion-specific effects that are important for the behavior of ions in soft matter.
\begin{acknowledgments}
  MLS, PV and IB acknowledge support from NSF-CHE. NL acknowledges support from a UIC University Fellowship and the GAANN program. ChemMatCARS is supported by NSF-CHE, NSF-DMR, and the DOE-BES. The APS at Argonne National Laboratory is supported by the DOE-BES. 
    \end{acknowledgments} 
\bibliography{bibliotheque}

\end{document}